# How Does Facebook Retain Segregated Friendship?
# An Agent-Based Model Approach

Firman M. Firmansyah

Department of Technology and Society, Stony Brook University

## Author Note

Corresponding concerning this article should be addressed to Firman M. Firmansyah, Department of Technology of Society, 1432 Computer Science, Stony Brook University, Stony Brook, NY 11794-4404. Email: manda.firmansyah@stonybrook.edu. Twitter: @firmansm.

This study was self-funded. The authors declare that they have no known competing financial interests or personal relationships that could have appeared to influence the work reported in this paper.

HOW DOES FACEBOOK RETAIN SEGREGATED FRIENDSHIP? 2**Abstract**

Facebook, the largest social networking site in the world, has overcome the structural barriers that historically constrain individuals to reach out to different others. Through the platform, people from all walks of life and virtually any location can develop diverse friendships online. However, friendships on Facebook have been as segregated as friendships in real life. This research sought to understand why the leading social networking site intended to 'bring the world closer together' retains segregated friendship. In doing so, we employed a series of agent-based simulations based on the Framework for Intergroup Relations and Multiple Affiliations Networks (FIRMAN). As demonstrated, Facebook has primarily enhanced users' capacity to maintain a larger number of friendships (tie capacity), while has done little to empower users' ability to accept diversity and befriend different others (tie outreachability). Facebook must focus on the latter should they truly wish to contribute to the development of a more inclusive society. While in this study we focus on ethnically segregated friendship on Facebook, we argue that the same explanation may also hold for racially and ideologically segregated friendships on other bi-directional social networking sites.

**How Does Facebook Retain Segregated Friendship? An Agent-Based Model Approach**

Diverse friendships have been shown to have a variety of benefits for individuals, including enhanced creativity (Lee et al., 2012), improved health outcomes (Pachucki & Leal, 2020), increased tolerance (Ikeda & Richey, 2009), and advanced career paths (Aten et al., 2017; Seebruck & Savage, 2014). However, most friendships tend to be homogeneous and segregated, as demonstrated by prominent examples in gender (Laniado et al., 2016), race (Campigotto et al., 2021), and ethnicity (Smith et al., 2014; van Tubergen, 2015). Historically, this phenomenon was driven by structural factors such as dispersed living location (Mouw & Entwisle, 2006), unbalanced demographic composition (Joyner & Kao, 2000), and divided foci (Feld, 1981), which constrains individuals to reach out to different others. Of course, psychological factors such as preference (Yu & Xie, 2017; Ilmarinen et al. 2016) and intergroup attitudes (Bahns et al. 2015; Fischer 2011) also play roles.

Nowadays, the invention of social networking sites such as Facebook has overcome the structural barriers in unprecedented ways. Through Facebook, people from all walks of life and virtually any location can connect and form friendships online (Bouchillon, 2021). For instance, a Pacific Islander in Hawaii can become Facebook friends with a European American in Maine; an African American in Mississippi can befriend Asian American in California on Facebook. Thus, it makes sense to assert that diverse friendship should be more prevalent or at least its rate increases in the online environment. It turns out, however, friendships on Facebook have been as segregated as friendships in real life (Wimmer & Lewis, 2010; Hofstra et al., 2017). Even more so, they have involved less salient dimensions such as personality traits (Lönnqvist & Itkonen, 2016), moral values (Dehghani et al., 2016), and political ideology (Huber & Malhotra, 2016).

Unfortunately, segregated friendship on social networking sites not only alienates the benefits of diverse friendship but also potentially magnifies the drawbacks of homogeneous friendship. For instance, it outranks minorities, depriving them of access to population resources (Karimi et al., 2018). Additionally, it can cause filter bubbles (Pariser, 2011) that make individuals get exposed to the same information multiple times. This condition may precipitate narrow-mindedness and further lead to an echo-chamber (Barberá et al., 2015; Boutyline & Willer, 2017), even violence. In Myanmar, for example, viral hate speech against Rohingya refugees circulating among Facebook friends arguably made Burmese users show a lack of empathy toward the ethnic minority (Fink, 2018). In the United States, former President Trump's posts regarding 2020 election results liked and shared by a large number of Facebook users, arguably contributed to the attack on the U.S. Capitol in early 2021 (Vukčević, 2021).

**Developing a New Friendship on Facebook**

Founded in February 2004, Facebook has been the largest social networking site in the world (Tankovska, 2021). As of 2021, Facebook allows users to send friend requests virtually to anyone available in any part of the world and maintain up to 5,000 friendship connections. The users who receive the friendship invitations can accept, delete, or ignore the requests. The sender will only get notified if their friend requests are accepted. They will never know if the recipients delete or ignore the friend request. Once connected as friends, Facebook users can do various things including but not limited to seeing shared personal information, sharing photos and videos, commenting on each other's posts, and sending direct messages. Indeed, the sharing features will vary depending on the user's privacy settings.

Past surveys have reported that 31% of American teenagers had Facebook friends who they had never met in-person (Lenhart & Madden, 2007). Another experimental research has revealed that 35% of Facebook users did accept friend requests from strangers even if they were bots (Boshmaf et al., 2011). From the receiver's perspective, users decided to develop a new friendship with someone on Facebook if they knew him or her in real life, shared similar interests, or had mutual friends (Rashtian et al., 2014). Moreover, compared to female users, male users tend to add more new friends on Facebook (Muscanell & Guadagno, 2012). From the sender's perspective, friend requests sent by more attractive or opposite sex users tend to get accepted (Patil, 2012). Ironically, this making-new-friends phenomenon does not lead to a more diverse social network as previously mentioned. Rather, it amplifies and extends segregated friendship, from offline to online (Wimmer & Lewis, 2010; Hofstra et al, 2017).

**Current Study**

This research sought to understand why Facebook, a leading social networking platform that has been initially designed to 'bring the world closer together' (Facebook, 2019), retains segregated friendship particularly along ethnic lines. In doing so, this study employed agent-based modeling derived from the Framework for Intergroup Relations and Multiple Affiliations Networks (FIRMAN) introduced by Firmansyah and Pratama (2021). We chose FIRMAN since it simplifies the otherwise very complex process of friendship development in a formal language. Moreover, its premise that friendship must be reciprocated in order to develop aligns with the friending mechanism on Facebook. In this regard, two users will not become friends on Facebook unless one accepts the other's friendship invitation. This agent-based simulation approach also allowed us to further investigate, which part of this extended friendship segregation has been nurtured by the platform, and which part comes naturally from the individuals. By recognizing these factors, such interventions to catalyze diverse friendship development on Facebook and other social networking sites in general can be further developed. In the following section, we briefly explain FIRMAN and its components (i.e., social identity space, social identity distance, length of ties, tie outreachability, and tie capacity).

**Framework for Intergroup Relations and Multiple Affiliations Networks (FIRMAN)**

FIRMAN (Firmansyah & Pratama, 2021) is a formal theoretical framework in which individuals or agents are represented as nodes (with specified coordinates) and their relationships as ties (with specified lengths) in *social identity space*. A social identity space is an abstract space constituted by *n* numbers of social identities derived from the groups to which the individuals belong (e.g., gender, ethnicity, sports clubs). FIRMAN has been useful to explain why homogeneous friendships persist even in a diverse place, where the probability of developing heterogeneous friendships is actually greater than a chance alone.

**Social Identity Distance and Length of Ties**

Consider six individuals $i \in \{A, \ldots, F\}$ of two different ethnic groups in a hypothetical population. Let color depict their social identity (SI) with orange for the ethnic majority group and purple for the ethnic minority group. The majority group is arbitrarily assigned to 0, while the ethnic minority group is assigned to 1. The one-dimensional social identity space for this population is depicted in Figure 1.

**Figure 1**
*A One-Dimensional Identity Space*

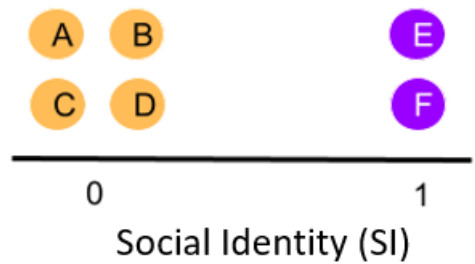

*Note.* Individuals of the same group are located in the same coordinate (SI = 0 for ethnic majority group and SI = 1 for ethnic minority group), and otherwise for individuals of different groups.

With the individuals located in the social identity space, we can easily calculate their social identity distance (w_dist) representing similarity/dissimilarity between any pair of nodes using Equation 1, which is derived from Manhattan distance (Black, 2019). Social identity distance differs from Borgadus' social distance, the latter of which is used to assess individuals' subjective acceptance of outgroup members as alters (Firmansyah & Pratama, 2021).

**Equation 1**
*Social Identity Distance*
$$w\_dist_{xy} = w \cdot \Sigma \, | SI_x - SI_y |$$
*Note.* $w\_dist_{xy}$ denotes social identity distance between individuals *x* and *y*; *w* denotes weight for social identity; *SI* denotes social identities.

In this example, $w\_dist_{xy} = 0$ means that nodes x and y are of the same group (e.g., A and C; E and F in Figure 1), while $w\_dist_{xy} = 1$ means otherwise (e.g., B and E; C and F in Figure 1). It is worth noting that the operation $\Sigma$ and constant *w* are particularly useful should the numbers of social identities measured in the study be more than two (Firmansyah & Pratama, 2021). As such, they represent that in fact not all social identities are equal. For instance, in the United States, people of different political affiliations are more prejudiced and thus are more distant toward each other than people of different races holding the other social identity constant (Iyengar & Westwood, 2015).

Such social identity distance will determine the length of ties (LT) needed to establish relationships between pairs of nodes in the social identity space. As shown in Figure 2, reciprocated friendships (undirected ties) between nodes of the same ethnicity (e.g., A and C; E and F) requires zero length of ties (w_dist$_{AC}$ = LT$_{AC}$ = 0; w_dist$_{EF}$ = LT$_{EF}$ = 0). Whereas friendships between nodes of different ethnicities (i.e., B and E) requires one length of ties (w_dist$_{BE}$ = LT$_{BE}$ = 1). It is worth noting that the zero length of ties (LT = 0) between two nodes are depicted by short tie rather than no tie (between a pair of stacked nodes) for practical and aesthetic reasons.

**Figure 2**
*Friendship Networks in a One-Dimensional Identity Space*

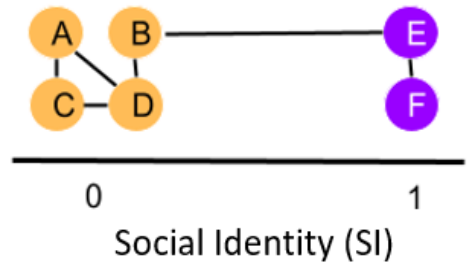

*Note.* Homogeneous friendship (e.g., A and C) requires zero length of ties (LT = 0) while heterogeneous friendship (e.g., B and E) requires one length of ties (LT = 1).

**Tie Outreachability and Tie Capacity**

FIRMAN (Firmansyah & Pratama, 2021) further postulates that nodes have different latent abilities concerning the maximum lengths and numbers of ties that they can generate in social identity space called *tie outreachability* (TO) and *tie capacity* (TC), respectively. Tie outreachability arguably represents the extent to which one can tolerate and accept group differences in real life. Whereas tie capacity arguably represents the number of friends one can keep at a time. Individuals cannot generate ties longer than their tie outreachability and more than their tie capacity. Thus, in social identity space, a reciprocated friendship between a pair of nodes will develop, *if and only if*, both can generate ties equal to their social identity distance, and both still have a capacity to do so. Mathematically, these conditions can be expressed as follows.

**Equation 2**
*Reciprocated Friendship Formation*

$$\text{friends}_{xy} \leftrightarrow \text{w\_dist}_{xy} \leq \min(TO_x, TO_y) \land \text{num\_f}_x < TC_x \land \text{num\_f}_y < TC_y$$

*Note. friends$_{xy}$* denotes reciprocated friendship between individuals *x* and *y*; *w_dist$_{xy}$* denotes weighted social identity distance between x and y; *num_f* denotes the number of friends each node currently has.

To put Equation 2 into context, let Table 1 define tie outreachability and tie capacity for individuals {A, ..., F}. All dyadic relationships in the social identity space presented in Figure 2 meet the conditions as stated in Equation (2). For instance, B and E are friends since their ties can reach out to each other positions (w\_dist$_{BE}$ = 1; TO$_B$ = 1; TO$_E$ =1) and their current numbers of friend are *not* more than their tie capacity (num\_f$_B$ = 2, TC$_B$ = 2; num\_f$_E$ = 2, TC$_E$ = 2).

**Table 1**

*Tie Outreachability and Tie Capacity of Individuals in Figure 2*

| Individual (Ego) | Tie Outreachability | Tie Capacity | Friends (Alters) |
|---|---|---|---|
| A | 0 | 3 | C, D |
| B | 1 | 2 | D, E |
| C | 0 | 2 | A, D |
| D | 1 | 3 | A, B, C |
| E | 1 | 2 | B, F |
| F | 1 | 3 | F |

*Note.* Friendship networks of these individuals are visualized in Figure 2.

It is worth highlighting that the social network in Figure 2 has reached equilibrium such that no more friendships can be developed. For instance, F could have been friends with A since their current numbers of friends are fewer than their tie capacity (num\_f$_F$ = 1, TC$_F$ = 3; num\_f$_A$ = 2, TC$_A$ = 3). However, they could not do so since A cannot generate ties equal to the social identity distance between them (TO$_A$ = 0; TO$_F$ = 1; w\_dist$_{AF}$ = 1). In another case, F could have been friends with D since their ties can reach out to each other's position (TO$_F$ = 1; TO$_D$ = 1; w\_dist$_{FD}$ = 1). However, since D has already reached their maximum tie capacity, they cannot generate another tie to develop a new relationship including with F (TC$_D$ = 3, num\_f$_D$ = 3; TC$_F$ = 3, num\_f$_F$ = 1). Indeed, the observed length and numbers of ties in a social identity space do not always the same with the nodes' tie outreachability and tie capacity.

**Hypothesis**

In light of FIRMAN, we argue that Facebook as a technology platform has enhanced individuals' tie capacity but *not* their tie outreachability. If in real life individuals can maintain only a few friendships, on Facebook they can keep in touch with even up to five thousand friends. Simultaneously, the platform has dramatically increased the pool of potential friends: from limited people nearby to limitless users worldwide. However, Facebook has little to no bearing on tie outreachability. In this respect, those who are intolerant offline remain intolerant online and still cannot accept group differences (Chaudhry & Gruzd, 2020; Matamoros-Fernández, 2017). These conditions arguably make Facebook retain segregated friendships. In other words, Facebook friendship patterns closely resemble those of offline friendships.

H$_1$: Friendships on Facebook are as segregated as friendships offline.

H$_2$: Facebook retains segregated friendship by increasing users' tie capacity but *not* their tie outreachability.

In the following, we will test these assertions using agent-based simulation. The results will be validated using empirical findings of a study examining Facebook users' offline and online personal networks in the Netherlands (Hofstra et al., 2017). We had particular interest in the referenced study since it directly examined users offline and Facebook friendships.

## Methods

As mentioned earlier, we used agent-based modeling (Macal, 2016) to test our hypotheses. This approach is particularly useful to make sense of macro-level social phenomena through micro-level behavioral detail (Squazzoni et al., 2014) and infer causal relationships between the independent and dependent variables that are otherwise difficult to be manipulated and captured in real settings due to ethical, economical, and practical reasons (Bonabeau, 2002; Jackson et al., 2017).

**Simulation Design**

We considered probability distributions proposed by Firmansyah and Pratama (2021) to generate agent populations in our simulation study. In this regard, we used the Bernoulli distribution to generate agents' social identity (SI) with *p* = .22. This parameter value is our effort to match the proportion of the majority-minority population in the referenced study used to validate the results. It should be noted, for parsimony purposes, we only incorporate one minority group, unlike the referenced study (Hofstra et al., 2017), which has five minority groups. We used social identity weight *w* = 1 since we only incorporated one social identity in this study.

Furthermore, we also used the Bernoulli distribution to generate tie outreachability (TO) with parameters $q \in \{.2, .5, .8\}$, representing populations with few, moderate, and many tolerant agents who can generate up to one length of ties. In the original manuscript, Firmansyah and Pratama (2021) proposed the binomial distribution to generate tie outreachability with parameters *m* (numbers of trial) and *q* (probability to succeed). Since our study only incorporates one social identity, we chose to employ the Bernoulli distribution, which is the special case of binomial distribution with one variable. We intentionally choose different notation *q* to distinguish it with SI's parameter *p* albeit derived from the same distribution.

To generate the tie capacity (TC), we used the normal distribution with $\mu \in \{3, 30\}$ and the same $\sigma^2 = .25$. We employed specific codes so that it only generates discrete numbers (see the source codes for review). These two parameters represent individuals' tie capacity in offline and Facebook environments, respectively. We generated agents n = 30 for the offline setting and n = 300 for the Facebook setting. As previously explained, Facebook increases the pool size of potential friends. The summary of the parameters used to generate agents is presented in Table 2.

**Table 2**

*Parameters to Generate Agent Populations*

| Case | Scenario | Parameters | | | | Simulation Trials |
|---|---|---|---|---|---|---|
| | | SI | N | TO | TC | |
| 1 | Offline | $p = .22$ | $n = 30$ | $q = .2$ | $\mu = 3, \sigma^2 = .25$ | 100 |
| | Facebook | $p = .22$ | $n = 300$ | $q = .2$ | $\mu = 30, \sigma^2 = .25$ | 100 |
| 2 | Offline | $p = .22$ | $n = 30$ | $q = .5$ | $\mu = 3, \sigma^2 = .25$ | 100 |
| | Facebook | $p = .22$ | $n = 300$ | $q = .5$ | $\mu = 30, \sigma^2 = .25$ | 100 |
| 3 | Offline | $p = .22$ | $n = 30$ | $q = .8$ | $\mu = 3, \sigma^2 = .25$ | 100 |
| | Facebook | $p = .22$ | $n = 300$ | $q = .8$ | $\mu = 30, \sigma^2 = .25$ | 100 |

*Note.* SI = social identity; N = number of agents, TO = tie outreachability; TC = tie capacity. Each case represents the comparison between offline and Facebook friendships under different conditions of TO parameters.

We call simulations sharing the same tie outreachability (TO) parameter as a *case* and simulations sharing the same tie capacity (TC) parameters as a *scenario*. For each scenario in each case, we performed 100 trials constituting 600 simulations in total. All simulations were conducted in R 4.1.0 with the help of RStudio 1.4.1103. The source codes and data sets are available as the supplementary materials.

**Simulation Procedure**

In all simulations, agents, for various reasons, need friends, and thus aim to develop friendships. They can only befriend other agents from their own population. The order of who gets to initiate and to whom is determined through a stochastic process. Agents who receive a friendship invitation cannot decline unless constrained by their tie outreachability or tie capacity. In this respect, they cannot befriend others whose social identity distance (w_dist) is greater than their tie outreachability and cannot develop any more friendship should their numbers of friends (num_f) have reached their tie capacity. This process will continue until no friendship can be further developed among agents in the population. In other words, the system has reached its equilibrium. Figure 3 illustrates the algorithms employed in this study.

**Figure 3**

*Friendship Development Algorithm*

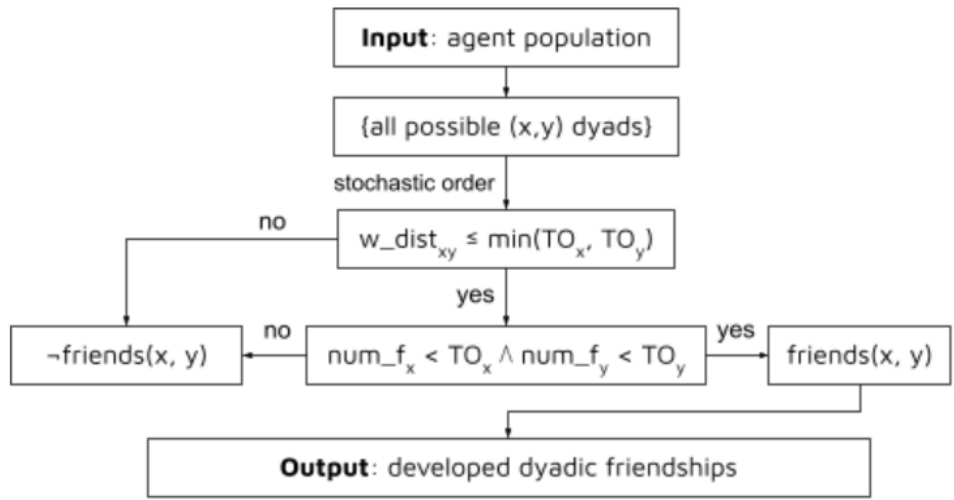

*Note.* Friendship will only develop between agents whose tie can reach out to each other's position in the social identity space and still have capacity to generate a tie. For each simulation, friendship development process does not stop until no more friendship can be developed.

**Analytic Strategy**

As for the analysis, *first*, we investigated if the developed friendships are ethnically segregated at both group and individual levels. At the group level, we examined the total numbers of homogeneous and heterogeneous friendship dyads in the population. At the individual level, we measured the percentage of ethnically similar alters (friends) for each ego (agent). *Second*, we compared offline scenarios with n = 30 and Facebook scenarios with n = 300 for each case and investigated if they share the same segregated patterns confirming our hypothesis. As previously mentioned, the results would be compared to the results of the referenced study (Hofstra et al., 2017). Finally, we also measured agent satisfaction by examining the agent's number of friends and comparing them with a threshold following Firmansyah and Pratama (2021). In this respect, agents would be satisfied should they have friends at least *half* of their tie capacity and be not satisfied, otherwise. While providing additional feedback to the system, these figures might help explain ways in which Facebook has intervened in users' friendships.

## Results

**Generated Agent Populations**

The proportions of ethnic majority and minority groups in all agent populations are depicted in Figure 4. On average, the ethnic majority group made up 78.02% of the population. This number is close to the ethnic majority percentage in the referenced study, which was 78.40%. As previously stated, we incorporated only one ethnic minority group for parsimony reasons.

**Figure 4**

*Generated Agents' Social Identity in All Simulations*

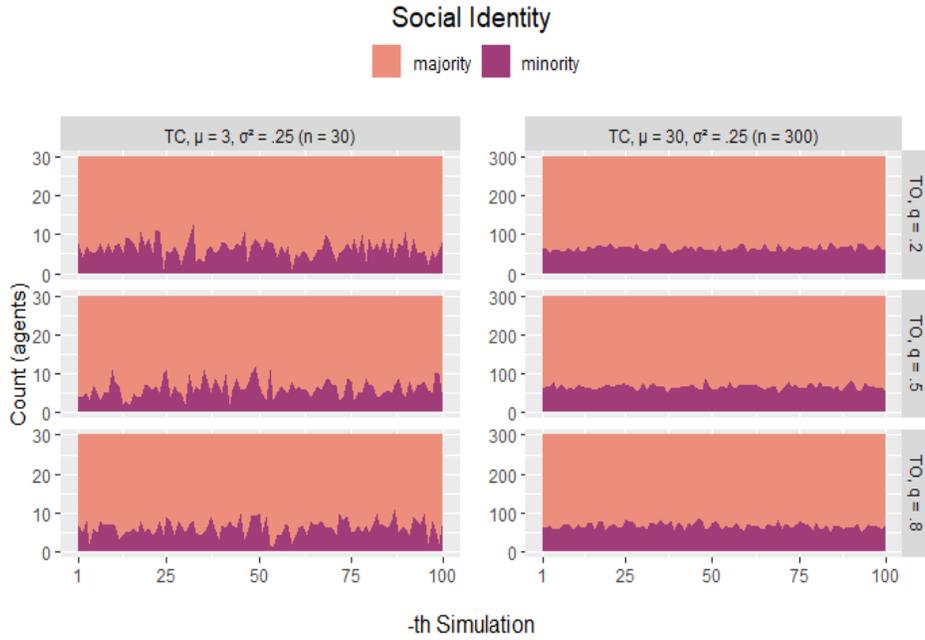

*Note.* The proportions of the ethnic majority (78.02%) and minority (21,98%) are relatively the same in all simulation scenarios.

**Figure 5**

*Generated Agents' Latent Variables in All Simulations*

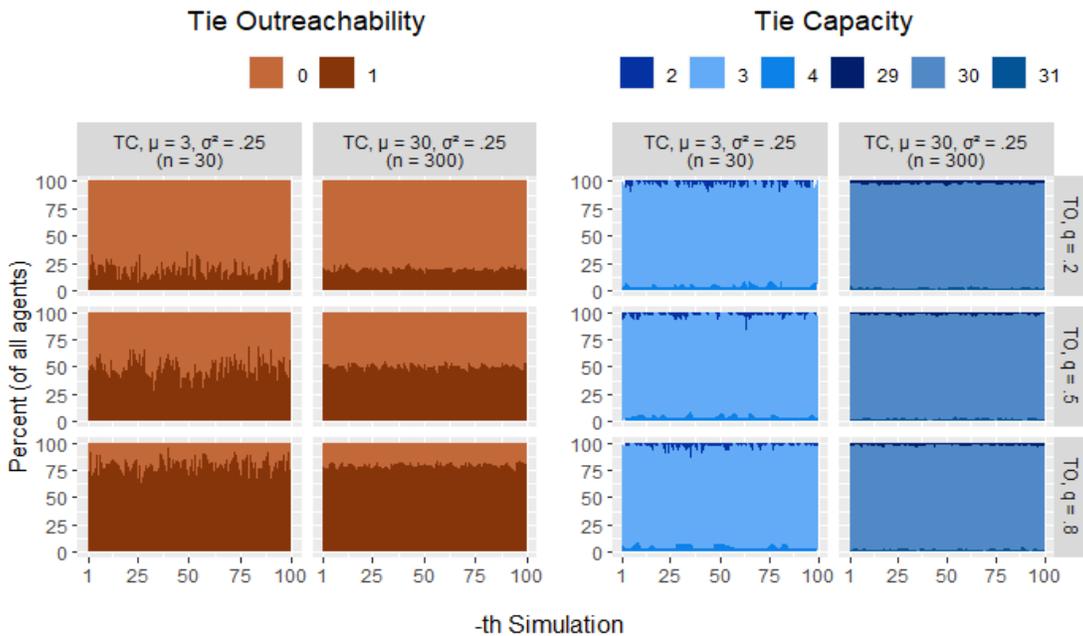

*Note.* The proportion of agents with tie outreachability (TO) ∈ {0, 1} are relatively equal across scenarios in the same case (horizontal comparison). The proportion of agents with tie capacity (TC) ∈ {3, 30} are relatively the same across cases with the same scenario (vertical comparison).

Figure 5 illustrates the agents' tie outreachability and tie capacity generated throughout the simulations. As can be seen, their variability is similar across scenarios and cases in which their parameters were held constant. For instance, the proportions of agents with TO = 1 are relatively the same across scenarios in the same case (horizontal comparisons). Whereas the proportion of agents with TC ∈ {3, 30} are relatively the same across cases with the same scenario (vertical comparisons).

**Developed Friendships**

Figure 6 visualizes a randomly chosen developed friendship network for each scenario and case. It should be noted that these networks are visualized using the Fruchterman-Reingold algorithm with the help of the igraph package (Csardi & Nepusz, 2006), not in FIRMAN's social identity space. Hence, the nodes' position and the length of ties connecting them in the network do not represent FIRMAN's social identity distance nor length of ties. At first glance, the developed social networks do appear to exhibit segregated friendship along ethnic lines, with the strongest group segregation occurring in Facebook scenarios where TO, $q$ =.3.

**Figure 6**

*Some Visualizations of Developed Friendship Networks*

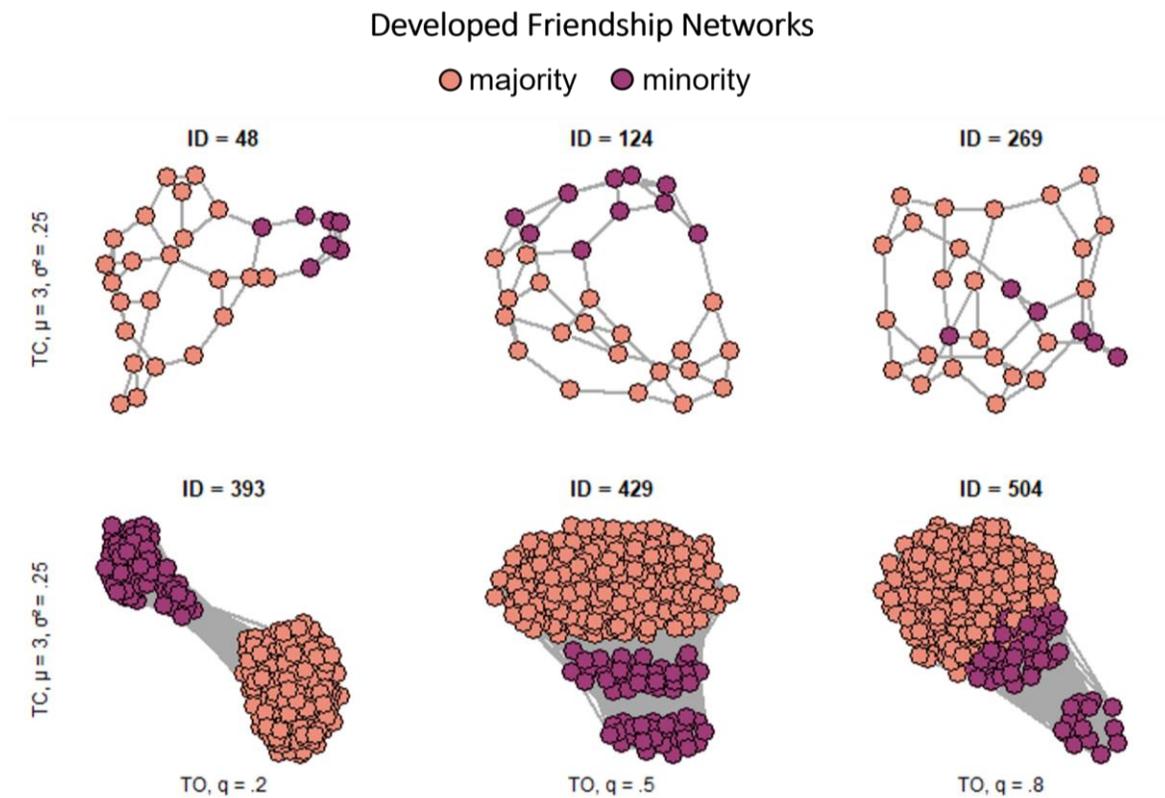

*Note.* Developed friendship networks visualized using Fruchterman-Reingold algorithm show segregated patterns.

Furthermore, the number of developed dyadic friendships in all simulations is depicted in Figure 7. As illustrated, the majority of friendships are formed between nodes of the same ethnicity. In contrast, heterogeneous dyads accounted for a relatively small proportion of total friendships, with the proportion increasing as the tie outreachability (TO) parameter values increased (vertical comparison). Marginal means of heterogeneous friendship dyads for TO, $q \in \{.2, .5, .9\}$ are 2.03%, 10.75%, 23.47% of the total dyads, consecutively. Meanwhile, the proportions of homogeneous friendships remain relatively constant across tie capacity (TC) parameters, despite the fact that the average number of total dyads increases dramatically from marginal means = 44 dyads for TC, $\mu = 3$ (offline scenario) to marginal means = 4,479 dyads for TC, $\mu = 30$ (Facebook scenario).

**Figure 7**

*The Number of Developed Dyadic Friendships in All Simulations*

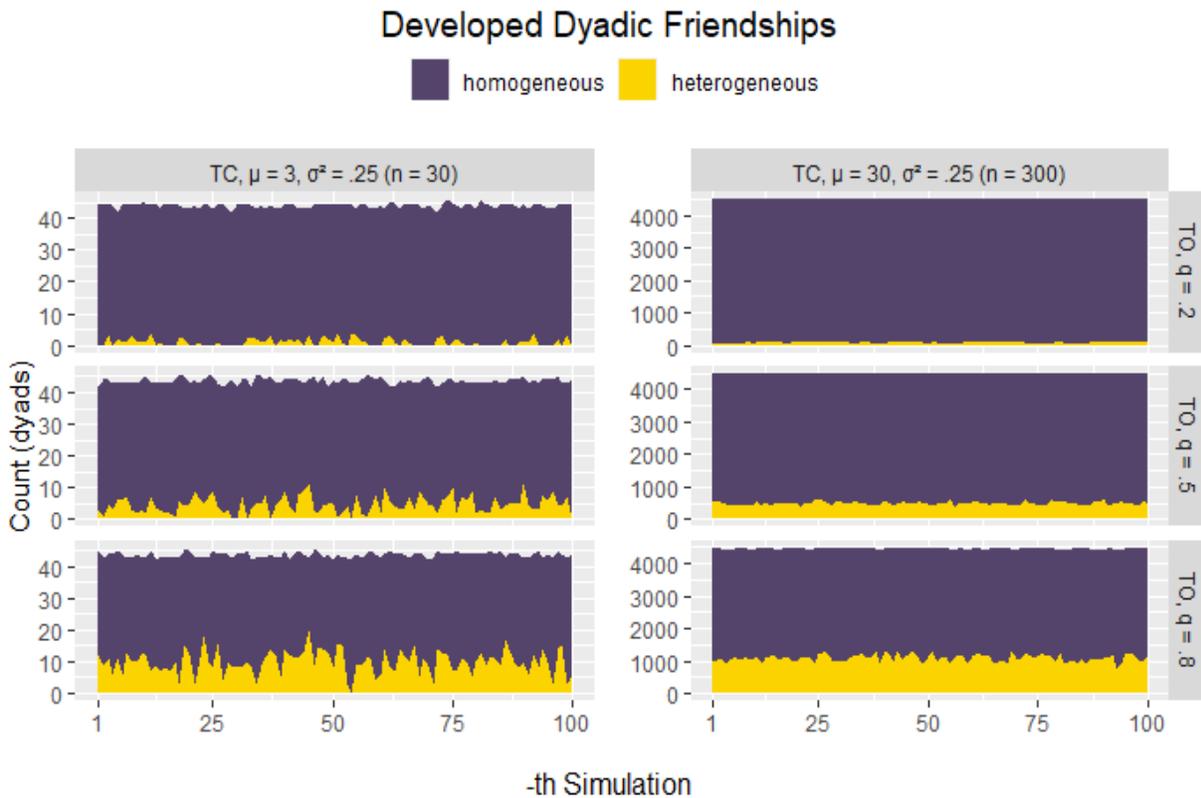

*Note.* Increasing the tie capacity (TC) parameter and population size (n), while holding the tie outreachability (TO) parameter constant does not lead to more heterogeneous friendship formations.

Figure 8 presents the percentage of alters (friends) who share the same ethnicity with the egos. As can be seen, both majority and minority members tend to befriend other agents of the same group. These numbers even exceeded the expected percentage of ego-alter similarities caused solely by random chance given the proportions of the majority and minority in the population, which are 78.02% for the majority and 21.98% for the minority. Moreover, it is interesting to note that simulations in the offline scenario have more variability compared to the simulation in the Facebook scenario.

**Figure 8**

*The Percentage of Ego-Alter Similarity in All Simulations*

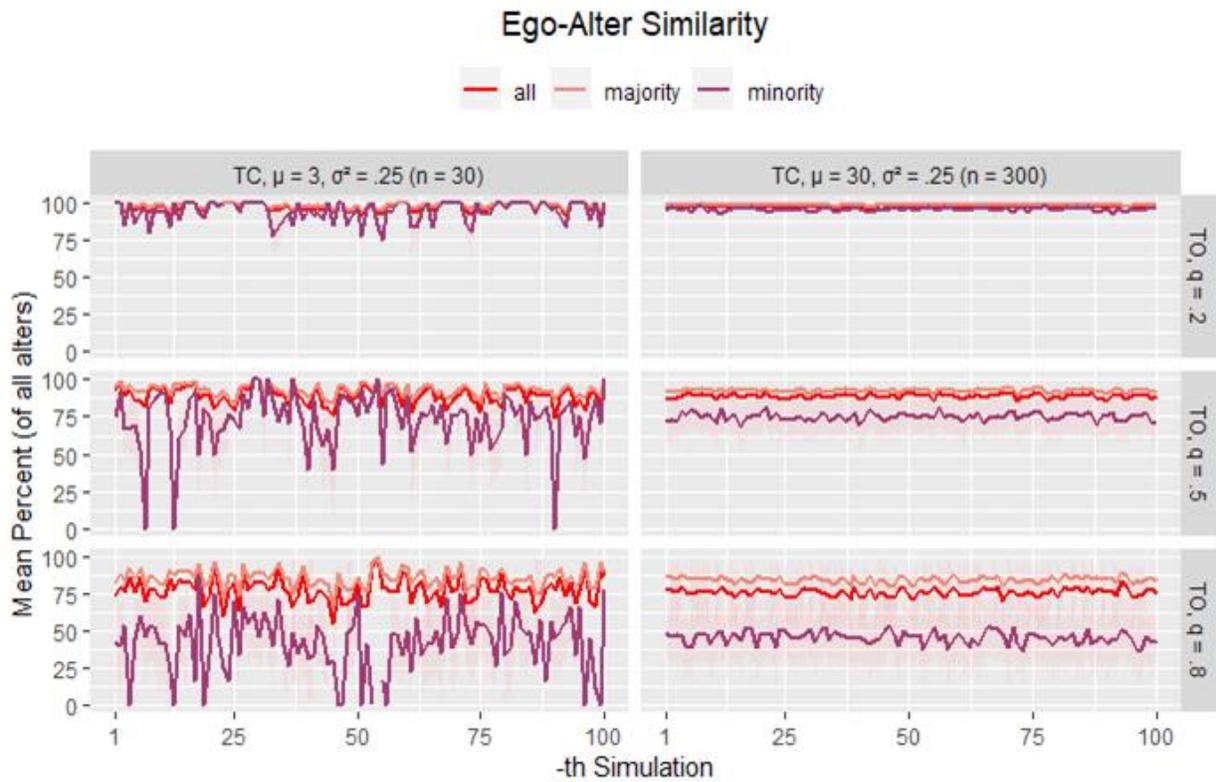

*Note.* The percentage of ego-alter similarity in Facebook scenarios (right column) seem to mirror in offline friendship scenarios (left column).

As mentioned earlier, we compared and contrasted the results with the findings of the referenced study investigating the same phenomenon. We only did this for the overall groups combined since in our study we only have one minority, whereas in the referenced study, there were five groups of minorities. In the referenced study (Hofstra et al., 2017), on average individuals had 76.218 % ethnically similar alters for offline scenario (friends in general in their term) and 75.974 % ethnically similar alters for Facebook scenario (Facebook friends without kinship in their term). As can be seen in Table 3, the third case has the most similar results with the referenced study.

**Table 3**

*Mean Percentage of Ego-Alter Similarity in Offline and Facebook friendships in All Simulations*

| Case | Scenario | Parameters | | Mean (SD) Percent |
| --- | --- | --- | --- | --- |
| | | TO | TC | |
| 1 | Offline | $q = .2$ | $\mu = 3, \sigma^2 = .25$ | 97.784 (7.073) |
| | Facebook | $q = .2$ | $\mu = 30, \sigma^2 = .25$ | 98.185 (0.959) |
| 2 | Offline | $q = .5$ | $\mu = 3, \sigma^2 = .25$ | 89.628 (7.853) |
| | Facebook | $q = .5$ | $\mu = 30, \sigma^2 = .25$ | 89.044 (1.020) |
| 3 | Offline | $q = .8$ | $\mu = 3, \sigma^2 = .25$ | 77.799 (5.575) |
| | Facebook | $q = .8$ | $\mu = 30, \sigma^2 = .25$ | 75.981 (0.651) |

To determine whether Facebook friendships replicate offline friendships, we divide the percentage of ego-alter similarities in the Facebook scenario by the percentage of ego-alter similarities in the offline scenario. We also did the same for the findings in the reference study. Table 4 presents the results.

**Table 4**

*The Ratio of Mean Percentage of Ego-Alter Similarity in Facebook and Offline Friendships*

| Case | Parameter | Facebook-Offline Friendships Ratio | | |
| --- | --- | --- | --- | --- |
| | TO | Simulation | Hofstra et al., 2017 | Δ |
| 1 | $q = .3$ | 1.004 | | -0.007 |
| 2 | $q = .5$ | 0.993 | 0.997 | 0.004 |
| 3 | $q = .8$ | 0.977 | | 0.020 |

*Note.* Δ = Facebook-offline friendship ratio in the simulation - Facebook-offline friendship ratio in the referenced study.

As can be seen, the ratios are almost identical in all cases. Only when the TO, $q = 3$, the ratios show different directions. In this respect, the proportions of ego-alter similarities are higher in the Facebook scenario than they are in the offline scenario.

**Agent Satisfaction**

The percentage of satisfied agents across all simulations is depicted in Figure 9. As can be seen, only in the offline friendship scenarios, did the simulations made some agents not satisfied. In contrast, in the Facebook friendship scenarios, the simulations made all agents have friends at least half of their tie capacity (TC) and thus satisfied.

**Figure 9**

*The Percentage of Agent Satisfaction in All Simulations*

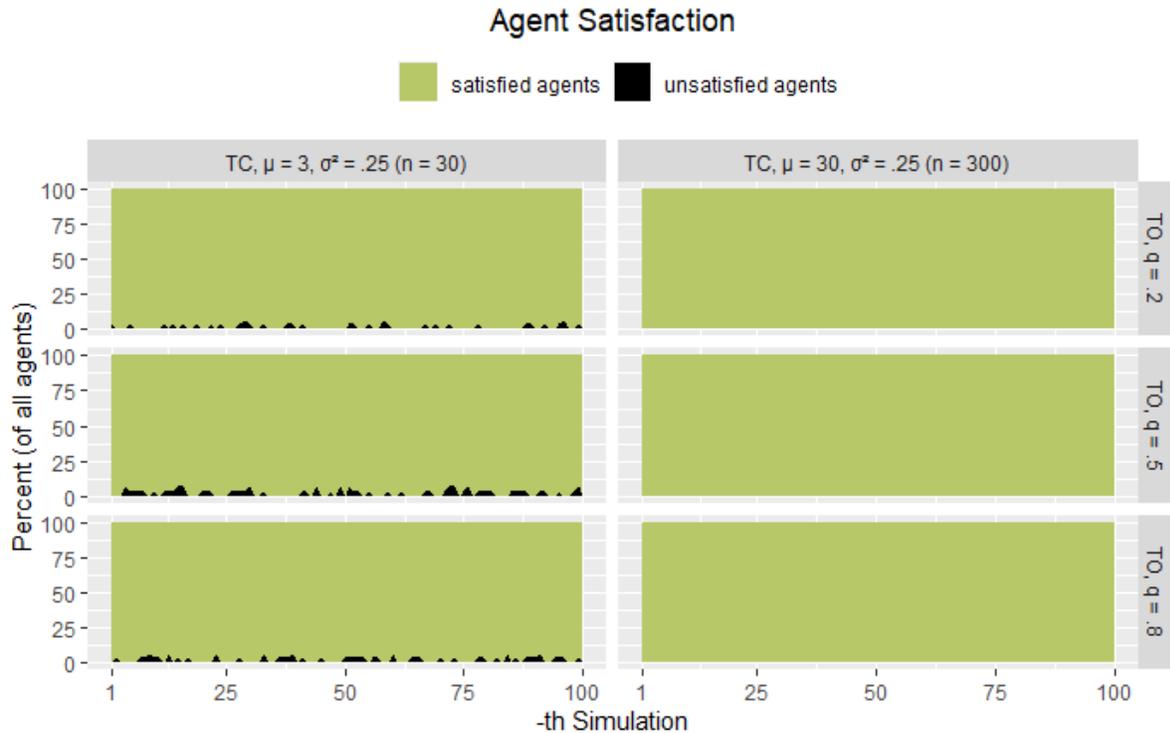

*Note.* Unsatisfied agents (having friends less than 50% of their tie capacity) are present in offline friendship scenarios only, albeit their percentage are considerably small.

## Discussion

The simulation results have confirmed our hypothesis that the patterns of Facebook friendship mirror the patterns of offline friendship, which are segregated along the ethnic lines. These findings corroborate previous empirical field study on the same subject (Hofstra et al., 2017; Wimmer & Lewis, 2010). While past research highlighting various antecedents including preference (Yu & Xie, 2017; Ilmarinen et al. 2016), living location (Mouw & Entwisle, 2006), demographic composition (Joyner & Kao, 2000), and foci structures (Feld, 1981), our study brings a different perspective explaining it through a basic individual level mechanism built on a formal framework, FIRMAN (Firmansyah & Pratama, 2021). In this respect, Facebook has increased users' tie capacity (TC) but has done little with users' tie outreachability (TO). Hence, the users, regardless of their level of tie outreachability, can keep in touch with a greater number of friends online. However, as mentioned earlier, those who are intolerant and racist offline remain intolerant and racist on Facebook (Chaudhry & Gruzd, 2020; Matamoros-Fernández, 2017), and thus cannot reach out to different others albeit they are on the very same platform.

Furthermore, our simulations demonstrate that even when the proportion of agents who are tolerant and thus capable of generating ties up to one length (LT = 1) is greater than 70% as depicted in Case 3, the percentages of homogeneous friendships remain higher than the percentages of heterogeneous friendships. In other words, both ethnic majority and minority members continue to gravitate toward similar others despite the fact that they neither despise nor prefer to affiliate with different others (actually we do not model individual preference in this simulation). This finding is consistent with Firmansyah and Pratama's argument (2021) that homogeneous networks are an inescapable social fact, even in a highly diverse and tolerant environment. This is because intolerant agents (TO = 0), who can only befriend similar others, prevent tolerant agents (TO = 1) of the same group from developing heterogeneous friendships with different others by reducing their tie capacity (TC). As a result, tolerant agents having homogeneous friendships are less capable of reciprocating, let alone initiating a new friendship with individuals of different groups.

Indeed, it is attempting then to point finger at the users, who cannot generate a longer tie to connect with ethnically different others on Facebook in the first place. However, if Facebook can empower users' tie capacity, why does it not, at least, try to enhance user tie outreachability as well? To our knowledge, Facebook has a history of rectifying unintended consequences of its products following the occurrence of terrible incidents that garner public attention (Fink , 2018; Vukčević, 2021). Should Facebook take the proactive step of empowering users' tie outreachability, the number of heterogeneous friendships will increase as shown in our simulations (comparison across cases). Even if heterogeneous friendships do not outnumber homogenous friendships, the percentage of users exposed to diverse perspectives by connecting with different others on the platform will still increase.

Additionally, Facebook appears to be more concerned with pleasing their users. As the findings show, all agents in the Facebook simulation scenario (TC = 30, n = 300) have been satisfied with their numbers of friends. Whereas in the offline scenario (TC = 3, n =30), some agents, even though they only made up less than 20 percent of the total population, have not been satisfied. In this regard, Facebook has helped users to deal with the constraints caused by the systems (e.g., limited tie capacity, dispersed geographical location) such as making agents of the same groups available as potential friends (either by empowering their tie capacity or increasing potential friend pool size). This argument can also explain why many users seem content with the status quo. They will be satisfied and happy as long as they can befriend other users and keep in touch with them on the platform. The fact that segregated friendships can alienate the benefits of diversity (Lee et al., 2012; Pachucki & Leal, 2020) and exacerbate the disadvantages of homogeneity (Karimi et al., 2018; Pariser, 2011) does not appear to bother the majority of them.

Finally, the simulations suggest what needs to be done to remedy the unintended consequence of the platform design, which retains friendship segregation instead of bringing diverse people closer together. As previously mentioned, Facebook must increase users' tie outreachability in the same way it increases users' tie capacity by improving their tolerance and acceptance of differences. Encouraging users to do fact checking (Miškolci et al., 2018) and banning users who perpetuate conflicts (Vukčević, 2021) are not the only way to do so. In fact, Facebook with its resources, among other things, can develop friendship algorithms that favor diverse friendship formation such as suggesting tolerant users to befriend different others instead of suggesting them to befriend those of similar interests or having mutual friends (who usually happen to be of the same ethnicity). This way, we believe, will expose intolerant users who happen to befriend tolerant users, toward a more diverse perspective, just like the notion of how extended intergroup friendship reduces prejudice (Yucel & Psaltis, 2017; Žeželj et al., 2017). Of course, this intervention, albeit aiming for a greater good, needs to be run ethically, for instance by asking for users' consent in advance.

**Study Limitations and Directions for Future Research**

This study has explained the extended segregated friendship phenomenon on Facebook that is otherwise complex in a parsimonious way through a series of agent-based simulations. With this strength, there are some limitations we need to consider. First, we have not yet taken into account the strength of the relationships on Facebook. We treat them as if they are similar. As past studies show (McMillan, 2022), there might be a different proportion of group homogeneity between strong ties (i.e., close friends) and weak ties (i.e., acquaintance). Second, in our simulations we have yet to include algorithms that possibly used by Facebook in intervening friendship formation such as friend suggestions based on mutual friends, similar interests, and same geographical locations. Employing this algorithm may yield even higher homogeneous friendship formations. We suggest that future research consider these limitations when designing future simulation studies on friendship formation on Facebook.

**Conclusion**

Facebook's mission 'brings the world closer together' as a social networking site platform seems to be a failure. Friendships on Facebook have been as segregated as offline friendships. Our study demonstrates that this is because Facebook primarily increases users' capacity to befriend more people, while doing little to improve users' ability to befriend diverse others. Facebook must focus on the latter should Facebook truly wish to contribute to the development of a more inclusive society. While in this study we focus on ethnically segregated friendship on Facebook, we argue that the same explanation may also hold for racially and ideologically segregated friendships on other bi-directional social networking sites.

HOW DOES FACEBOOK RETAIN SEGREGATED FRIENDSHIP? 22Mouw, T., & Entwisle, B. (2006). Residential segregation and interracial friendship in schools. *American Journal of Sociology*, *112*(2), 394–441. https://doi.org/10.1086/506415

Muscanell, N. L., & Guadagno, R. E. (2012). Make new friends or keep the old: Gender and personality differences in social networking use. *Computers in Human Behavior*, *28*(1), 107–112. https://doi.org/10.1016/j.chb.2011.08.016

Pachucki, M. C., & Leal, D. F. (2020). Is having an educationally diverse social network good for health? *Network Science*, *8*(3), 418–444. https://doi.org/10.1017/nws.2020.14

Pariser, E. (2011). *The filter bubbles: What the internet is hiding from you*. Penguin UK.

Patil, S. (2012). Will you be my friend?: Responses to friendship requests from strangers. *ACM International Conference Proceeding Series*, 634–635. https://doi.org/10.1145/2132176.2132318

Rashtian, H., Boshmaf, Y., Jaferian, P., & Beznosov, K. (2014). To befriend or not? A model of friend request acceptance on Facebook. *Symposium on Usable Privacy and Security (SOUPS)*, 285–300. https://www.usenix.org/system/files/conference/soups2014/soups14-paper-rashtian.pdf

Seebruck, R., & Savage, S. V. (2014). The differential effects of racially homophilous sponsorship ties on job opportunities in an elite labor market: The case of NCAA basketball coaching. *Sociological Inquiry*, *84*(1), 75–101. https://doi.org/10.1111/soin.12021

Smith, J. A., McPherson, M., & Smith-Lovin, L. (2014). Social Distance in the United States: Sex, Race, Religion, Age, and Education Homophily among Confidants, 1985 to 2004. *American Sociological Review*, *79*(3), 432–456. https://doi.org/10.1177/0003122414531776

Squazzoni, F., Jager, W., & Edmonds, B. (2014). Social simulation in the social sciences: A brief overview. Social Science Computer Review, 33(3), 279–294. https://doi.org/10.1177/0894439313512975

Stecklow, S. (2018). Why Facebook is losing the war on hate speech in Myanmar. In *Reuters*. https://www.reuters.com/investigates/special-report/myanmar-facebook-hate/

Tankovska, H. (2021, February 9). *Global social networks ranked by number of users 2021*. Statista. https://www.statista.com/statistics/272014/global-social-networks-ranked-by-number-of-users/